\documentclass[aps,prl,reprint,english,onecolumn,superscriptaddress,amsmath,amssymb,floatfix]{revtex4-2}

\usepackage[utf8]{inputenc}
\usepackage[english]{babel}
\usepackage{amsmath,amsfonts,amssymb}
\usepackage[T1]{fontenc}
\usepackage{xurl}
\usepackage{soul}
\usepackage{changes}

\usepackage[linktoc=all,hidelinks]{hyperref} 

\hypersetup{
    colorlinks=true, 
    citecolor=blue,
    linktoc=all,    
    linkcolor=black,  
}
\usepackage[version=3]{mhchem}
\usepackage{stmaryrd}
\usepackage{epstopdf}
\usepackage{graphicx}
\graphicspath{{./Figures/}}

\begin{document}
\title{{Nonlinear Optical Encoding Enabled By Recurrent
Linear Scattering}}

\author{Fei Xia\footnote{Corresponding author: F.X. (\href{mailto:fei.xia.oe@gmail.com}{fei.xia.oe@gmail.com})}}
\affiliation{Laboratoire Kastler Brossel, ENS-Universite PSL, CNRS, Sorbonne Université, Collège de France, Paris, 75005, France.}

\author{Kyungduk Kim}
\affiliation{Department of Applied Physics, Yale University, New Haven, Connecticut 06520, USA}

\author{Yaniv Eliezer}
\affiliation{Department of Applied Physics, Yale University, New Haven, Connecticut 06520, USA}

\author{SeungYun Han}
\affiliation{Department of Applied Physics, Yale University, New Haven, Connecticut 06520, USA}

\author{Liam Shaughnessy}
\affiliation{Department of Applied Physics, Yale University, New Haven, Connecticut 06520, USA}

\author{Sylvain Gigan\footnote{Corresponding author: S.G. (\href{mailto:sylvain.gigan@lkb.ens.fr}{sylvain.gigan@lkb.ens.fr})}}
\affiliation{Laboratoire Kastler Brossel, ENS-Universite PSL, CNRS, Sorbonne Université, Collège de France, Paris, 75005, France.}

\author{Hui Cao\footnote{Corresponding author: H.C. (\href{mailto:hui.cao@yale.edu}{hui.cao@yale.edu})},}
\affiliation{Department of Applied Physics, Yale University, New Haven, Connecticut 06520, USA}

\begin{abstract}
     {Optical information processing and computing can potentially offer enhanced performance, scalability, and energy efficiency. However, achieving nonlinearity, a critical component of computation, remains challenging in the optical domain. Here, we introduce a design that leverages a multiple-scattering cavity to passively induce optical nonlinear random mapping with a continuous wave laser at a low power.} Each scattering event effectively mixes information from different areas of a spatial light modulator, resulting in a highly nonlinear mapping between the input data and the output pattern. We demonstrate that our design retains vital information even when the readout dimensionality is reduced—thereby enabling optical data compression. This capability allows our optical platforms to offer efficient optical information processing solutions across applications. We demonstrate our design's efficacy across tasks, including classification, image reconstruction, key-point detection, and object detection, all achieved through optical data compression combined with a digital decoder. In particular, high performance at extreme compression ratios is observed in real-time pedestrian detection. Our findings open pathways for novel algorithms and unconventional architectural designs for optical computing.
\end{abstract}
\maketitle

\section*{Introduction}

Optical information processing leverages the unique properties of light, such as its parallelism, which allows the simultaneous processing of multiple data streams, and low energy consumption \cite{shastri2018, kues2017, xu202111, wetzstein2020inference, shastri2021photonics}. Moreover, light possesses a vast frequency spectrum, enabling ultra-high bandwidth and data throughput \cite{shen2017deep, rotter2017light, wetzstein2020inference, shastri2021photonics}. By exploiting these characteristics, optical information processors have the potential to unlock new levels of performance, scalability, and energy efficiency, which could transform the landscape of information processing in the optical domain   \cite{wetzstein2020inference, shastri2021photonics}. It has enabled new applications when coupled with existing optical instruments, such as imaging system \cite{chang2018hybrid} to enhance the performance.

However, the full potential of optical processors can only be realized by overcoming certain challenges, one key requirement is the optical nonlinear mapping   \cite{shen2017deep, hughes2018training}. Nonlinear mapping is essential to approximate arbitrary function and has been a powerful element in neural networks as it allows the models to recognize complex patterns and approximate any given function   \cite{goodfellow2016deep}. It plays a vital role in representation and feature learning, as it facilitates the discovery of higher-level, more informative, and discriminative features for a task   \cite{lecun2015deep,wang2023image}. The application of nonlinear mappings allows the extraction of abstract and nonlinear features, thereby enhancing the input data representation   \cite{krizhevsky2012imagenet, lin2018all}. In existing optical computing platforms, optical nonlinear mapping has been primarily achieved using nonlinear optical materials, which provide a nonlinear relationship between input and output fields   \cite{tait2017neuromorphic, miller2015are,  wang2018chip, teugin2021scalable, williamson2019reprogrammable, li2022all, zhou2022nonlinear,shirdel2016photonic}. Optical nonlinearity often requires intense pumping and high peak power, which can be
energy demanding, necessitates design and engineering of nonlinear or active materials, and generally restricts to lower order nonlinear mapping with limited tunability.  \cite{shen2017deep, hughes2018training}. Alternatively, conversion of signals from optical to electrical and back to optical is used for nonlinear processing of optical data, but with limited speed. 

{Herein, we propose to exploit the passive nonlinear optical mapping inside a multiple-scattering cavity   \cite{eliezer2022exploiting}, akin to the steady state of a reservoir computer, for rapid optical information processing. High-order nonlinearity fosters the generation of low-dimensional latent feature space and facilitates strong data compression. Previously propagation through a multiple scattering material has been exploited to perform linear optical random projections  \cite{saade2016random}, followed by an intensity detection. It can be regarded as a single random layer neural network, and has been used for multiple machine learning tasks  \cite{rafayelyan2020large, dong2019optical, brossollet2021lighton, ohana2021photonic}, but remains limited in performances by its intrinsic linear behavior of the mapping. By introducing multiple scattering in a cavity design, we enable multiple bounces on the same input pattern, creating effectively a optical nonlinear transformation of the input data, without the need of nonlinear optical materials or optical-electrical-optical conversion typically used for nonlinearity in optical information processing. }We demonstrate high computing performances across tasks from classification, image reconstruction, key-point detection, and object detection, with the optically compressed output fed into a digital decoder. Notably, we show that our system exhibits high performance even at a mode compression ratio (defined by the input macro pixel numbers on DMD to output number of speckle grains on the camera) of $\sim$3000:1, for high-level computing tasks, as evidenced in real-time pedestrian detection with bounding box generation. Our work illuminates the role of varying nonlinear orders in optical data compression based on mutual information analysis, and paves the way for tunable optical nonlinear mapping and energy-efficient computing,

\section*{Results} 
\subsection{{Nonlinear random mapping with tunable nonlinearity}} 

Introducing nonlinearity has long been a challenge and simultaneously a necessity in optical computing platforms. Nonlinearity is a key element for enabling complex operations and boosting computational power   \cite{shastri2021photonics, wetzstein2020inference}. It is particularly important for approximating arbitrary functions - a task critical in machine learning. In this study, we present a novel approach to address this challenge by utilizing nonlinear mapping provided by multiple linear scattering of light within an optical cavity  \cite{eliezer2022exploiting}. We constructed the multiple-scattering cavity using an integrating sphere (Fig. 1a), which features rough inner surface that scatters light. A continuous-wave (CW) laser operating at low power is injected into the cavity via the first port, resulting in an output speckle pattern from the second port.  The third port integrates a Digital Micromirror Device (DMD) to display input patterns. In general, Born series can be used to describe the scattering process in the cavity:

\begin{equation}
E_{out} = \mathbf{T}E_{in} = \left[ \mathbf{V} + \mathbf{V}(\mathbf{G}_o\mathbf{V}) + \mathbf{V}(\mathbf{G}_o\mathbf{V})^2 + \ldots \right] E_{in}
\end{equation}

Here, matrix $\mathbf{T}$ represents a linear mapping from the input optical field $E_{in}$ to the cavity to the output field $E_{out}$. $\mathbf{V}$ is the matrix denotes the scattering potential inside the cavity, and $\mathbf{G_o}$ is the Green's matrix representing light propagation within the cavity in between bounces off the boundary. The notation $\mathbf{(G_oV)}^n$ represents the matrix $\mathbf{G_oV}$ multiplied by itself $n$ times. The final intensity image formed on the camera is given by $I_{cam}=|E_{out}|^2$, where the $|.|^2$ represents an element-wise operation. The $\mathbf{T}$ expansion begins with a term indicative of single scattering and subsequent terms indicate multiple scattering events in the cavity.  In the cases where a single scattering is the dominant event, the mapping from $\mathbf{V}$ to $E_{out}$ is predominantly linear. In our case with multiple scattering, the relation between the scattering potential configuration $\mathbf{V}$, and the output field $E_{out}$, becomes nonlinear. {The Born series can also be reformulated as:
${\bf T} = {\bf V} \sum_{m=1}^{\infty} ({\bf G_0} {\bf V})^{m-1} = {\bf V} \sum_{m=1}^{\infty} {\bf U} {\bf \Lambda}^{m-1} {\bf U}^{-1}$, where ${\bf G_0} {\bf V} = {\bf U} {\bf \Lambda} {\bf U}^{-1}$, ${\bf \Lambda}$ is a diagonal matrix of elements equal to eigenvalues of ${\bf G_0} {\bf V}$, the corresponding eigenvectors are columns of ${\bf U}$. For high order $m$, the largest eigenvalue $\lambda_{\rm max}$ dominates over all other eigenvalues, and the polynomial orders in ${\bf T}$ can be approximated as $\sum_{m=1}^{\infty} \lambda_{\rm max}^{m-1}$. Thus the nonlinear coefficient decays exponentially with order $m$. Experimentally due to chaotic ray dynamics in our cavity, it is difficult to extract the largest eigenvalues for different active areas on DMD.} Furthermore, since part of the surface area of the cavity can be modified by the DMD using the input (modulation) patterns, this also provides a reconfigurable scattering potential inside the cavity. Multiple bounces of light off the modulated area of the DMD results in a nonlinear mapping from the input pattern displayed on the DMD to the output speckle pattern. As the number of bounces on DMD increases, the order of this nonlinear mapping increases (Fig. 1b and 1c). It is this nonlinear relationship that forms the foundation for the passive nonlinear encoding technique that we explore in this work. 

\begin{figure*}[!htbp]
 \centering
  {\includegraphics[width=1\linewidth]{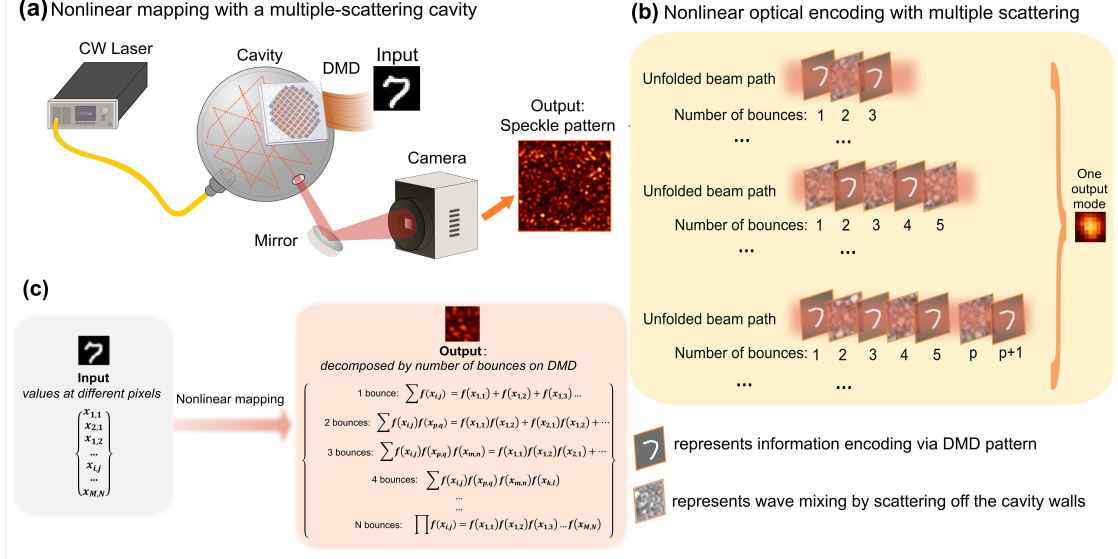}}
  \caption{Concept of using a multiple-scattering cavity as a passive, tunable nonlinear optical information processor: (a) The experimental setup, in which the key component for creating the passive nonlinear random mapping is a DMD mounted on an integrating sphere. The output of the cavity produces a fully developed speckle pattern, with its response being nonlinear in the geometric configuration of the DMD; (b) Representative figure showing the cavity essentially encodes the input pattern on the DMD through optically mixing different areas of input through multiple bounces to create a highly nonlinear feature -- a speckle recorded by a camera. (Input pattern is adapted from MNIST dataset \cite{lecun1998mnist}.) (c) The mathematical representation of a nonlinear mapping process that transforms a set of input elements on the DMD into a collection of nonlinear features in the output speckle pattern. Multiple scattering in the cavity generates mixed terms of the input valeus at different pixels with various high nonlinear orders, which provide rich nonlinear features that can be optimally trained to enhance performance in complex computational tasks. $f(x)$ denotes the operation of scaling the configuration of a DMD macropixel $x_{i,j}$. }
 \label{fig:1}
\end{figure*}

 Light scattering within the cavity can be adjusted solely by altering the pattern displayed on the DMD, as an input pattern (see Fig. 1a). Each micromirror on the DMD can be toggled between two positions. This action effectively modifies the scattering potential $\mathbf{V}$ for light, determining the mapping from the input pattern on the DMD to the output optical field $E_{out}$. {A larger modulation area boosts the probability of light scattering by the modulated part of the scattering potential, thereby enhancing the nonlinear mapping. The more times light is scattered by the DMD pattern, the more chance it samples the input pattern (Fig. 1b). Each scattering event effectively mixes the information from different parts of the DMD, resulting in a complex optical encoding of the entire pattern. The longer the light remains in the cavity, the more encoding and mixing occur, effectively ensuring that light in each output mode (speckle grain) carries information about a multitude of input data (Fig. 1b).} The interaction due to multiple scattering events results in a nonlinear mapping where the intensity of each output mode (speckle grain) becomes a highly nonlinear function of the input pattern (Fig. 1c). The number of bounces determines the order of nonlinearity. To further enhance the nonlinear order, the number of bounces is increased by covering the output port by a partial reflector to increase the dwell time of light inside the cavity. Such nonlinear mapping induced by multiple scattering is purely passive (no need for high power) and is fundamentally distinct from the traditional nonlinear optics that rely on intrinsic material response. . 

This scheme offers an efficient means of achieving tunable high-order nonlinear random mapping at constant low power with a continuous wave (CW) laser in a passive manner (see Methods), compared to conventional optical nonlinearities that rely on the material response at high optical intensity  \cite{agrawal2000nonlinear, boyd2008nonlinear}. In our case, the nonlinear order is independent of input power, and can be tuned rapidly ($\sim$20 kHz) by altering the DMD modulated area. This rapid tuning capability outperforms many known nonlinear effects, such as thermo-optical nonlinearity   \cite{wang2016thermo, ryou2021free}. Additionally, our scheme avoids dynamic chaos and instabilities commonly associated with conventional nonlinear optical systems and lasers   \cite{agrawal2000nonlinear, boyd2008nonlinear, ohtsubo2013}.

To comprehend and characterize the tunable nonlinear mapping introduced by our system, we explore how deep neural networks can function as a proxy to understand the nonlinear random mapping in our system. As detailed in the Supplementary Section 3, we find that the higher-order nonlinear mapping, provided by a larger area of modulation on DMD, can be approximated by a deeper neural network {(see also Methods part on the reformation of the Born series in terms of its proxy as deep neural networks with fixed random weights)}.

\subsection{Enhanced image classification}
To evaluate whether this nonlinear mapping can indeed provide any computational benefits, we begin by testing on a simple but widely recognized machine learning benchmark task, the FashionMNIST dataset  \cite{xiao2017fashion}. FashionMNIST is a popular fashion image classification challenge that includes 60,000 training samples and 10,000 test samples, each image measuring 28$\times$28 pixels.

We input the FashionMNIST data on the DMD and directly read the output speckle pattern to obtain both higher- and lower-dimensional representations. These representations, which we refer to as nonlinear features of the input information, can be utilized to execute computing tasks. To achieve nonlinear random mapping with tunable nonlinearity, given a dataset with fixed input size, we either adjust the modulated area of the DMD (Fig. 2a) or partially close the output port to change the number of times light is scattered by the DMD. During the training phase, we train only the linear digital layer using the nonlinear features generated from the training dataset at each given configuration. In the inference/test stage, we forward the output images from the cavity to the trained linear digital layer to generate predictions. (see the Methods part for details). 

\begin{figure*}[!htbp]
  \centering
  {\includegraphics[width=1\linewidth]{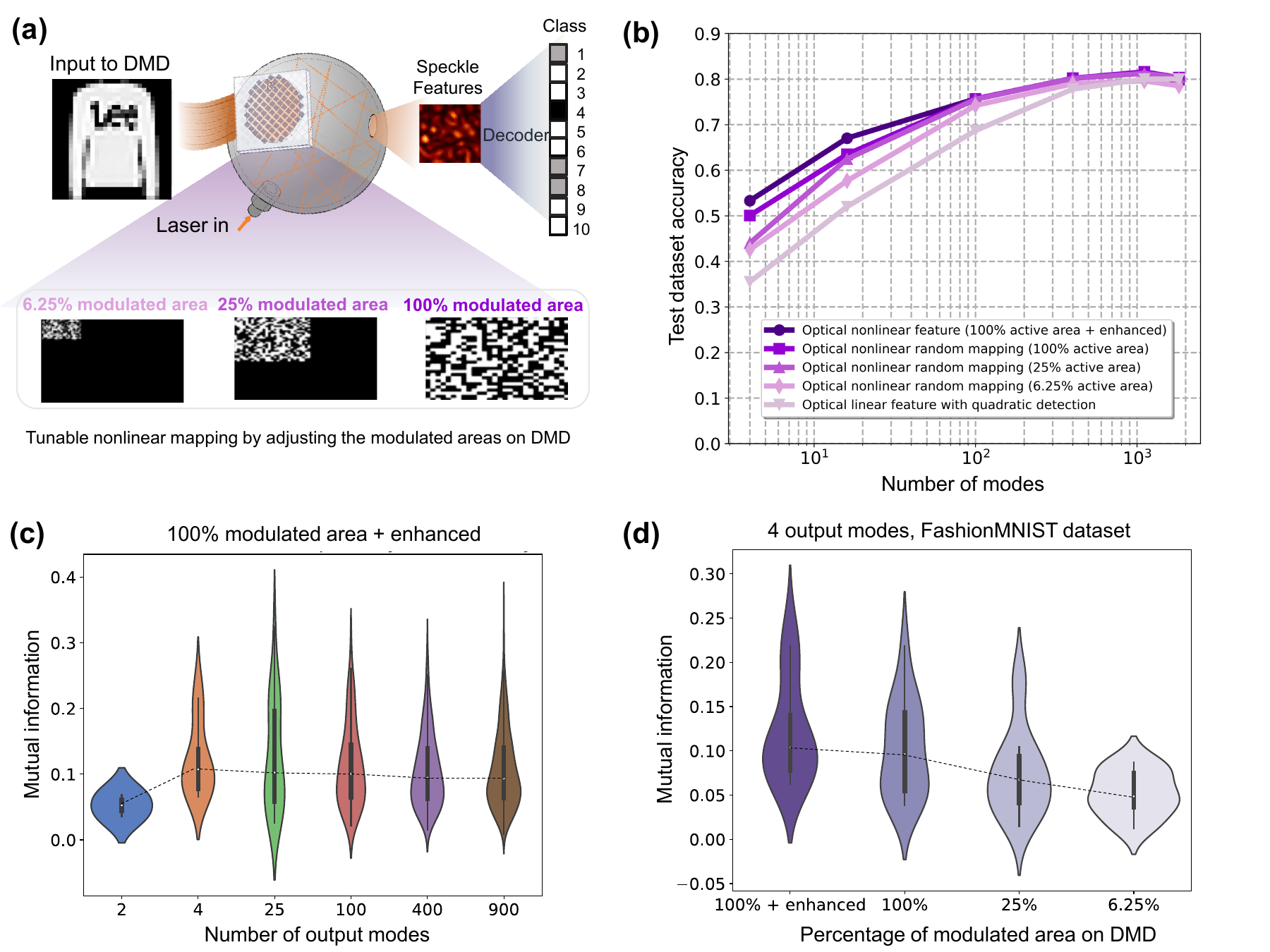}}

  \caption{Classification with nonlinear mapping. (a) Training data from the FashionMNIST datasets are used to train a 1-layer neural network as a digital decoder for classification tasks. Additionally, the percentage of the modulated area on the DMD is changed among 6.25$\%$, 25$\%$, 100$\%$ to adjust the order of nonlinear mapping. With full (100$\%$) modulation of DMD, nonlinear order is further enhanced by covering the output port with a partial reflector (silicon wafer). (b) FashionMNIST classification results with a linear classifier are presented under different numbers of output modes (speckle grains) and varying nonlinear strength. {The "optical linear features with quadratic detection" are simulated by scattering from a single layer with intensity detection to create a quadratic nonlinear response. Note that a linear regression for binarized Fashion-MNIST data cannot exceed 77.6\% with the same number of modes}. 
  (c,d) Violin plots representing the distributions of mutual information between the speckle grains and classification targets under varying numbers of output modes in (c), and differing order of nonlinear mapping by changing the modulated area on DMD or partially closing the cavity (enhanced) in (d). For $n$ speckle mode ($n$ on the $x$-axis), $4n$ replicated measurements from the same input were performed in (c) and (d). The dashed line plots depict the median values of the mutual information.
  Each violin's width reflects the distribution of the mutual information values of the speckle grains and its probability density. Within each violin: the slim black vertical line represents the range of minimal and maximal values; the black box represents the first to third percentile; the white dot represents the median.
  (c) Mutual information analysis when the number of output modes (speckle grains) varies under the highest-order nonlinear mapping. 
  (d) Mutual information analysis with low-dimensional speckle features (4 output modes) for FashionMNIST as a function of the nonlinear orders varied by modulated area on DMD, showing the advantage of going to higher-order nonlinear mapping.}
  \label{fig:2}
\end{figure*}

In Figure 2b, we present the classification performance in the FashionMNIST dataset using a linear classifier. 
To quantitatively compare the performance of different nonlinear strengths in the optical encoder, we fixed the linear decoder and used test accuracy as a metric for comparison. We observe that stronger nonlinearity leads to improved classification performance, particularly when the number of optical modes/speckles is smaller. This indicates that each speckle from higher-order nonlinear mapping embeds more information. These findings further suggest that our device may possess a unique advantage in optical data compression. 

To more comprehensively quantify the information within each spatial mode (speckle grain) in our output images, we employ the concept of mutual information. Compared to regression, mutual information includes both linear and nonlinear dependencies and does not make assumptions about the underlying data distribution  \cite{cover2012elements}. It is widely used in compressive sensing  \cite{donoho2006compressed}, a technique focused on efficiently acquiring and reconstructing sparse or compressible signals, and has found significant applications in machine learning for tasks such as feature selection  \cite{peng2005feature}, model interpretation  \cite{chen2016xgboost}, and understanding variable dependencies  \cite{kraskov2004estimating}. In our case, we calculate the mutual information between the output features and the target classes for the dataset (see details in Method and Supplementary Section 4). This quantifies how well the nonlinear optical features contain the abstract information that is useful for high-level computing tasks (see Supplementary Section 4 for details). In Fig. 2c and 2d. the violin plots effectively illustrate the distribution of mutual information between the speckle grains and the classification targets. A notable observation from the results is the onset of saturation of mutual information required for FashionMNIST classification with 4 to 25 modes/speckles, as shown in Fig. 2c. This saturation occurs under the highest-order nonlinear mapping in our experiments. Further, Fig. 2d underscores the benefit of escalating to higher-order nonlinearity. We observe that, indeed, higher-order nonlinear mapping creates stronger mutual information between the features and the targeted classes, given the same number of output modes/speckles. This observation implies that our system can more effectively capture the underlying relationships between the features and the target classes when higher-order nonlinear mapping is introduced.

\subsection{Demonstration with complex tasks}
\subsubsection{Image reconstruction }

Building on the enhanced information provided by the nonlinear features from our system, we pose the question: can this enhanced information (within a few output modes) yield superior image reconstruction? To address this, we conduct a comparative analysis of the nonlinear features generated in two distinct scenarios: one featuring a higher order nonlinear optical random mapping induced in the multiple-scattering cavity (Fig. 3c), and another that presents a linear optical random projection (Fig. 3a)   \cite{saade2016random} with nonlinearity only at the detection stage (intensity measurement).

To most efficiently extract the embedded information from a few speckles, we deviate from the traditional approach of employing a digital linear layer for classification and, instead, introduce a customized and optimized multilayer perceptron (MLP) as a decoder for image reconstruction. The architecture of this decoder is fine-tuned using neural architecture search to optimize image reconstruction. Subsequently, we train two digital decoders, each featuring optimal architectures, on the FashionMNIST dataset under high compression ratios of $\sim$ 31:1 (only $\sim$ 25 modes), as illustrated in Fig. 3.

\begin{figure*}[!htbp]
  \centering
  {\includegraphics[width=0.98\linewidth]{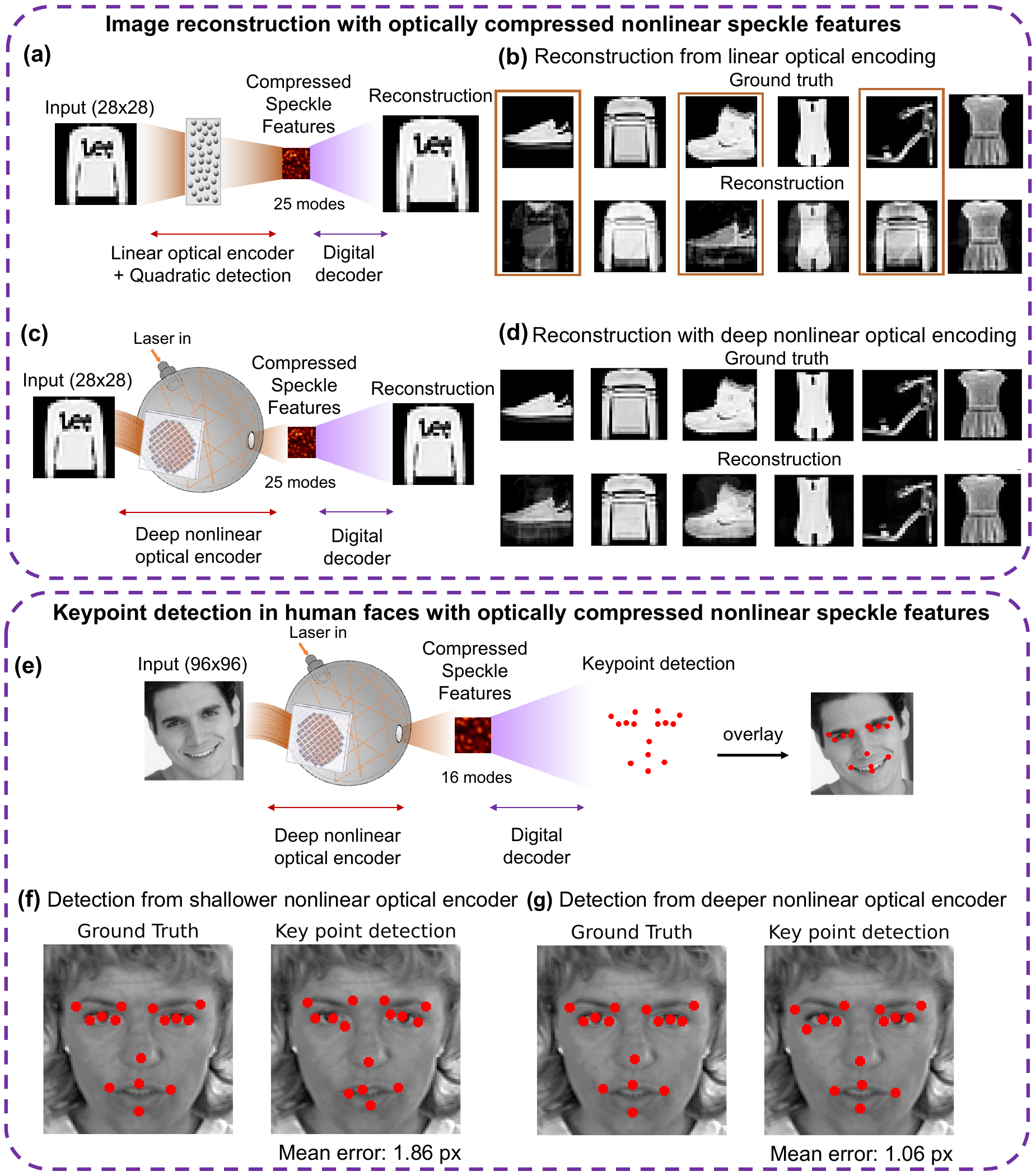}}
 
  \caption{Computing performance enhanced by nonlinear optical data compression: Concept of image reconstruction using (a) linear optical complex media for linear encoding and camera detection with quadratic response; (b) reconstruction from speckle features from (a); orange boxes represent the wrong reconstructed pairs; (c) the multiple-scattering cavity as a nonlinear optical encoder also with camera detection and employing compressed speckle features for digital reconstruction of the original image data. (d) Reconstruction from speckle features generated by the multiple-scattering cavity. In (b,d), approximately 25 speckle grains are used with a compression ratio of 31:1 and are used to train two digital decoders (see details in Methods). It is demonstrated that, given the same number of compressed output modes (speckle grains), nonlinear features generated from the cavity can provide a reduced mean squared error by {0.6}, resulting in a better reconstruction of the images in (d) compared to (b). {More results included in the Supplementary Figs. S4-6}. (e) Concept of keypoint detection in human faces (images with 96 $\times$ 96 pixels) with compressed speckle features (f) keypoint detection with mode compression ratio of 576:1, using 16 output modes with relatively weaker nonlinearity (25$\%$ modulated areas in DMD) and a 5-layer MLP decoder (g) improved keypoint detection with a reduced mean error in pixels across 15 keypoints (1.06 pixels compared to 1.86 pixels errors in (f)), using 16 output modes (speckle grains) with relatively stronger nonlinearity (full modulated areas in DMD) and an 9-layer MLP decoder.}
  \label{fig:3}
\end{figure*}

It is of note that, despite the optimally trained decoder in each case, the quality of the reconstructed images varies (Fig. 3b and 3d, more in Fig. S4 and S5). {We observe that augmented nonlinear random mapping indeed facilitates improved image reconstruction with the mean squared error of $\sim$1.4 in the test set (Fig.3d, Supplementary Fig. S5) compared with that of $\sim$2.0 in linear optical features (Fig. 3b, Supplementary Fig. S4), each under a separately optimized decoder architecture. When using the same decoder architecture, the nonlinear features still outperform with the mean squared error of $\sim$1.5 (Supplementary Fig. S6).}

Our findings show that the nonlinear optical mapping in our system can efficiently compress and retain vital information while also decreasing data dimensionality. Motivated by these results, we are prompted to explore the potential of nonlinear features in executing other high-level computing tasks.

\subsubsection{Keypoint detection}

A key advantage emerging from our work is that optical data compression, facilitated by multiple scattering in the cavity, generates mixtures of highly nonlinear features. These are particularly useful for applications that require high-speed analysis and responses of high-dimensional data. Our DMD contains 4 million pixels and can accommodate large images. However, in our image reconstruction demonstration, the input dimension of the FashionMNIST dataset is limited to 28 $\times$ 28, creating an inherent upper limit for the maximum compression ratio that can be demonstrated. A major strength of our system is its ability to easily scale up both the size of the input data and the effective neural network's depth of the optical encoder without increasing input power, thus allowing for an efficient representation of the input information in an energy efficient way. This adaptability and scalability facilitates tackling more complex tasks and processing larger datasets without losing crucial information.

Pushing the compression further and exploring other high-level computing tasks, we delve into two specific applications where we scale up the input images. A significant example, as depicted in Fig. 3e, demonstrates that we can extract 15 key points from human face images  \cite{facehuman} with an order of magnitude improvement in the mean squared error, which has decreased from 0.208 (using 25$\%$ modulated area in DMD) to 0.014 (using 100$\%$ modulated area in DMD enhanced with the partial reflector), due to the incorporation of stronger nonlinearity with larger modulation area, even when the number of output modes (speckle grains) is reduced to 16 (Fig. 3f-g). {In both cases, the architectures of the decoders are separately optimized and trained for optimal performance. Even when we use the same architecture (a 5-layer MLP) that was optimized for features from a 25\% modulated area in DMD for the decoder to train features from the latter case, the mean squared error associated with these higher nonlinear features remains low ($\sim$ 0.3).} This task, crucial for various applications such as facial recognition, emotion detection, and other human-computer interaction systems, illustrates the robustness of our approach in dealing with high-level tasks while maintaining a high compression ratio.  {An additional advantage of our methodology lies in its implications for privacy protection and adversarial robustness, as our method can securely encode facial information in random speckle grains.} 

\subsubsection{Real-time video analytics}

\begin{figure*}[!htbp]
 \centering
  {\includegraphics[width=1\linewidth]{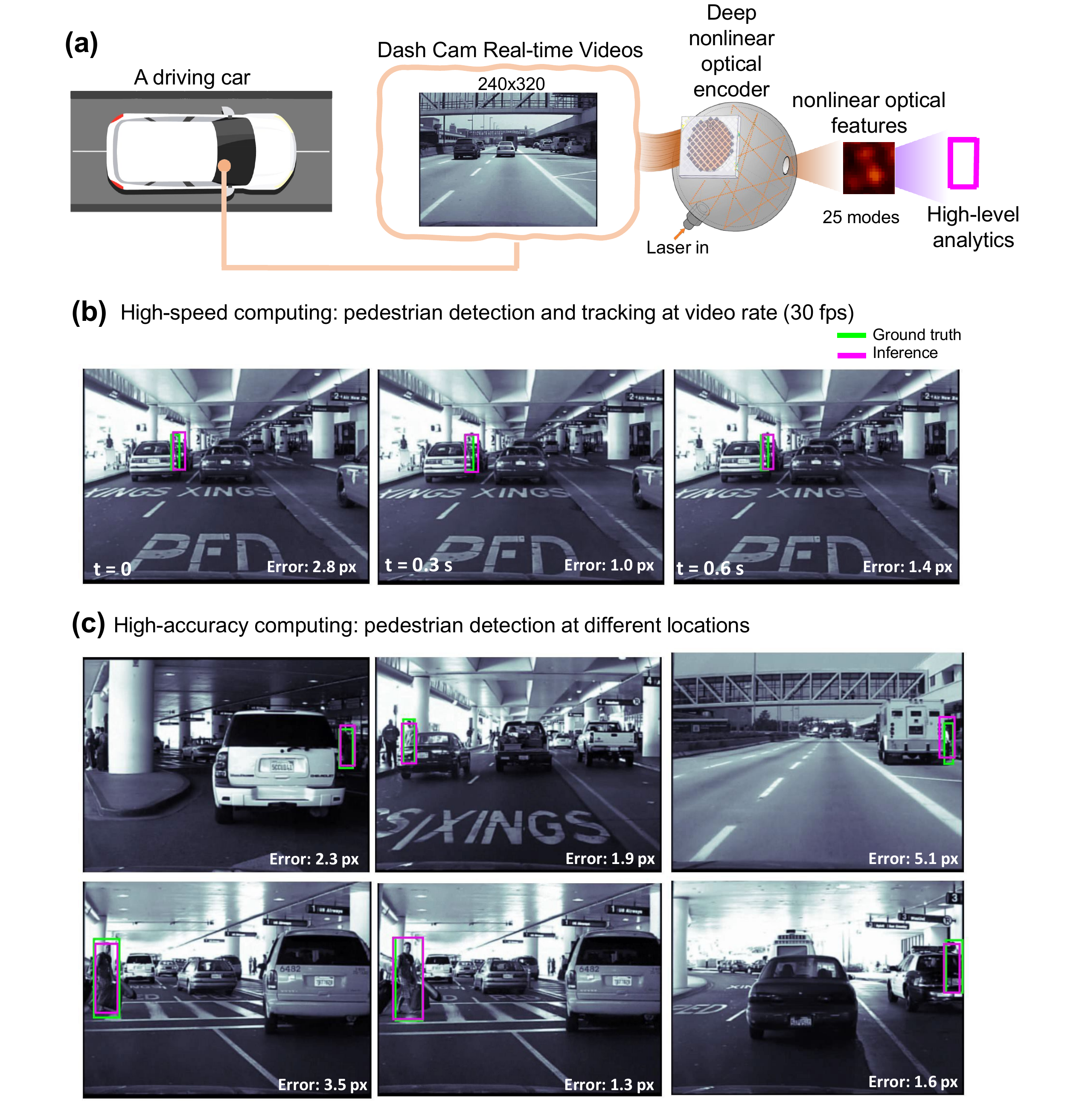}}
  \caption{Real-time video pedestrian detection in driving with high mode compression ratio using only 25 output modes: (a) Schematic representation of real-time pedestrian detection using video data from a dash camera during driving. The multiple-scattering cavity functions as an optical data compressor, and compressed nonlinear optical features are utilized for pedestrian detection with a digital decoder. (b) Demonstration of pedestrian detection at {close to} a real-time video rate. The magenta boxes represent the inference results from the speckle. The green boxes represent the ground truth. The speed of optical processing, that is, nonlinear feature generation, is as fast as light, and its readout speed is limited by only the camera. With only 25 modes, our camera can currently reach at least 800 Hz. The inference time with the 25 modes in pedestrian detection is 0.0035 s, leading to a total response time (inference + generation of optical features) of less than 0.1 s, which is faster than the typical human response time of 0.2$\sim$22 s. The error unit is in pixels (px). (c) Demonstration of pedestrian detection at various locations during continuous video streaming; the mean detection error with only 25 modes remains within 1.92 pixels (px).}
 \label{fig:4}
\end{figure*}

The last application we demonstrate is real-time video analytics, using the benchmark dataset known as the Caltech Pedestrian  \cite{dollar2009pedestrian}, including real-time video recordings (Fig. 4a). The images from the videos displayed on DMD have a size of 240$\times$320 pixels (see Methods for details). Using our multiple-scattering cavity, we can compress the data to achieve a compression ratio of up to 3072:1 (i.e., using only 25 output modes), while maintaining high positional accuracy (Fig. 4b) within mean squared errors of 1.92 pixels in identifying pedestrian positions at high speed $\--$ 0.0035 seconds in total response time (including compressed optical nonlinear feature generation and inference time) with an optimized digital backend (a 10-layer MLP) per frame (Fig. 4c, Supplementary Movies 1 and 2).

This application is particularly critical in the field of autonomous vehicles and advanced driver-assistance systems, where high-speed pedestrian detection is essential to ensure safety and allow fast reaction time. The high compression ratio of our system, combined with its rapid processing speed, shows great promise for such applications where fast and accurate detections are imperative.

To further estimate the gain we have in terms of optical data compression, we calculated the number of parameters and operations in the digital domain with and without an optical encoder. In human face keypoint detection: Our method with optical encoder demonstrated a mean pixel error of 1.06, slightly surpassing the performance of a widely-utilized conventional CNN architecture (which is still widely used as a benchmark for vision tasks). With a CNN model (1 conv + 1 pooling + 1 conv + 3 fully-connected layer), we achieved a mean pixel error of approximately 1.23. The digital CNN model comprises over 74 million parameters and requiring around 83 million operations, are 2 orders of magnitude higher than the digital operations/parameters used in our system (approximately 310k digital trainable parameters/operations). This comparison underlines the enhanced accuracy of our approach, while also pointing to a significant reduction in computational complexity and resource utilization inherent to our method. For pedestrian tracking, our method exhibited mean pixel errors ranging from 1.3 to 3.6, closely matching the performance of a conventional CNN model used for comparison, with which we obtained mean pixel errors between 1.37 and 3.33. The comparison model in this instance incorporates 3 convolutional layers and 2 fully-connected layers, involving more than 39 million parameters and necessitating approximately 172 million operations. In contrast, our decoder after non-linear optical projection used only about 1 million parameters and a similar number of operations (1 million) to achieve comparable performance. In general, the higher the input dimension and the more we compress, there more digital operations we can allocate into optical domain therefore better leverage the advantage of information processing of light.

\vspace{0.1cm}

\section*{Discussion} 
{Exploiting optics for computing, which brings benefits such as high speed, large bandwidth, and parallelization, has traditionally been impeded by the challenge of addressing optical nonlinearity. Conventional all-optical methods typically involve complex experimental conditions,  using nonlinear materials, like nonlinear crystals or polymers, pumped by high-power short-pulsed lasers, or semiconductor lasers operating in continuous or pulsed modes  \cite{skalli2022photonic, van2017advances}}. Although these have shown optical computing benefits in a variety of platforms including multimode fiber  \cite{teugin2021scalable}, integrated photonics  \cite{li2022all}, and free-space optics, limitations regarding their robustness, energy efficiency, and stability persist.

In this work, we completely avoid the limitations of conventional optical nonlinearity by proposing a unique approach to achieve optical nonlinear random mapping by utilizing multiple scattering within an optical cavity. This strategy enables us to institute nonlinear random mapping, where the adjustment of nonlinearity is entirely dependent on the geometrical configuration and the Q factor of the cavity, thereby influencing the scattering potential. The intrinsic mixing of input information within the dataset at varying nonlinear orders permits us to generate highly nonlinear features compared with traditional optical nonlinear mappings, especially those with solely lower-order (2 to 3) nonlinearity that most nonlinear materials practically permit. From a machine learning perspective, by expanding to higher-order nonlinear mapping, we essentially generate an augmented optical feature space, incorporating more mixtures of higher-level input information. This expansion increases mutual information between the subspace of the feature space and the entire input pattern (evident by the image reconstruction task) and output targets (Fig.2, and other high-level tasks), facilitating a higher compression ratio for complex tasks. In essence, our system demonstrates the capability to execute optical data compression by harnessing multiple scattering of light in a reconfigurable cavity. This approach allows for the efficient preservation of critical information while also stringently reducing data dimensionality. 

We have demonstrated that our multiple-scattering cavity, equipped with passive and tunable nonlinear optical random mapping capabilities, can act as an optical nonlinear encoder with adjustable nonlinearity. Our system can deliver enhanced computing performance in a low-dimensional latent feature space for a range of computing tasks, from image classification to higher-level tasks such as image reconstruction, keypoint detection, and object detection when trained with a lightweight digital backend. This approach might offer considerable benefits for high-speed analytics in both scientific and real-world applications. Our system permits easy scaling of both the input data and the effective depth of the neural networks approximating the optical encoders, providing an efficient optical representation of large-scale input patterns using a limited number of output modes. This versatility helps to manage more intricate tasks and process larger data sets without substantial loss of vital information.

{Our nonlinear mapping system functions as a reservoir computer in a steady state. This is also the case of other systems that have been realized before  \cite{boikov2023direct, porte2021complete, teugin2021scalable} but comparatively, our design allows for easy scaling up and tuning of nonlinearity without varying the input power. Furthermore, our system may serve as a trainable physical neural network  \cite{wright2022deep}, if one part of the DMD is utilized for an input pattern and another is tuned or trained for direct readout without the need for digital processing.} The performance of our computing tasks can be further improved by, e.g., replacing the binary DMD with an analog SLM for information encoding. The detection part of our system can be further improved by replacing the camera with a fast photodetector array, given the small number of output modes that need to recorded for decoding. 

Our current optical computing architecture is beyond one-to-one architectural mapping of the digital neural network (DNN). It may inspire next-generation optical computing to exploit nonlinear mappings beyond conventional schemes and promoting the development of more energy-efficient neuromorphic computing platforms including  \cite{chen2023deep} and beyond optics \cite{momeni2023physics, del2018leveraging, momeni2023backpropagation}, where nonlinearity can be effectively harnessed and utilized.  Our findings could also spark new research directions in fields such as optical data compression for imaging  \cite{chen2015optical, weng2023non, li2021spectrally} and sensing  \cite{wang2023image, muminov2020fourier}, optical communication  \cite{wu2014secure}, and quantum computing  \cite{venkataraman2013phase}, where innovative nonlinear mechanisms can significantly enhance performance, efficiency {and potentially opportunity to enhancing data privacy and adversarial robustness  \cite{ohana2021photonic, bezzam2022privacy, cappelli2022adversarial}}.

\vspace{0.2in}
\noindent{During the final stage of this work, we became aware of the independent work of very different optical machine learning implementations that are based on the same principle of realizing nonlinear processing with linear optics   \cite{yildirim2023nonlinear, wanjura2023fully}.  }\\

\noindent \textbf{Acknowledgements}:
The research conducted at Yale University is partly funded by the US National Science Foundation (NSF) under Grant No. ECCS-1953959 and the US Air Force Office of Scientific Research (AFOSR) under Grant No. FA9550-21-1-0039. H.C. acknowledges the discussions with Prof. Ulrich Rührmair and Michael Lachner. S.G. and F.X. acknowledge funding from Chan Zuckerberg Initiative (2020-225346). F.X. acknowledges the funding from Optica Foundation Challenge Award 2023. The authors acknowledge Prof. Peter McMahon, Prof. Logan Wright, Prof. Tianyu Wang, Jianqi Hu, Jonathan Dong, and Zipei Nie for feedback on the initial manuscript. The authors acknowledge Jianqi Hu for careful proofreading of the manuscript. The authors acknowledge Bruno Loureiro for insightful discussions. \\

\noindent \textbf{Author contribution}:
F.X., S.G., and H.C. designed the research. F.X. further developed the concept and performed the research and prepared the manuscript with input from all authors, F. X. and S. H. performed the analysis. Y.E. built the experimental setup. K.K., Y.E., S. H., L.S. collected the experimental data. All authors discussed the results. K.K., Y.E., S.G., and H.C. edited the manuscript. S.G. and H.C. initiated and supervised the project.\\

\noindent \textbf{Competing Interests}: The authors declare no competing financial or non-financial interests.

\vspace{0.5cm}

\noindent\textbf{Methods}
\medskip
\begin{footnotesize}

\noindent {\textbf{Further explanation of Born series}:
To better connect the Born series with neural network, such as multi-layer perception (MLP), Born series can also be re-written as:
\begin{equation}
E_{out} = \mathbf{S}^n E_{in} = \mathbf{S}_u(\mathbf{S}_{u-1}(\ldots (\mathbf{S}_1(\mathbf{V}))\ldots)) E_{in}
\end{equation}  
 where \( \mathbf{S}_n(\cdot) \) represents a scattering operator, \( \mathbf{S}_1(\mathbf{V}) = \mathbf{V} \), for \( n > 1 \), \( \mathbf{S}_{n+1}(\mathbf{V}) = \mathbf{V} + \mathbf{S}_{n}(\mathbf{V})\mathbf{G}_0\mathbf{V} \). The iterative expression of the Born series involving the scattering operator \( \mathbf{S} \) can be seen to structurally resemble the iterative operation in a MLP, where data are transformed across multiple layers in an iterative way. However, $\mathbf{V}$ and $\mathbf{G}_0$ are identical in all layers.}
 \\

\noindent\textbf{Setup information}:
The multiple-scattering cavity is an integrating sphere with a rough inner surface and a diameter of 3.75 cm. The cavity has three ports on its boundary: one port is attached to a digital micromirror device (DMD, Texas Instruments DLP9000X) which provides a reconfigurable scattering potential, while the other two ports serve as the input and output ports of the cavity. A single-frequency continuous-wave laser (Agilent 81940A, wavelength = 1550 nm) at 21.3 mW is coupled through a single-mode fiber into the cavity through the input port. Upon multiple scattering events at the rough inner surface, the output light escapes the cavity via the output port. From the spectral correlation width of output speckle pattern, the average pathlength of light inside the cavity is estimated to be approximately 100 meter. The average number of bounces off the cavity boundary is on the order of 5000. To capture the output intensity pattern, a mirror is positioned adjacent to the output port, directing the output light towards an InGaAs camera (Xenics Xeva FPA-640). A linear polarizer is placed in front of the camera to record the speckle intensity patterns, which represent a complex nonlinear relationship between the configuration of the DMD and the resulting output speckle. \\

\noindent \textbf{Experimental procedure of computing with the multiple-scattering cavity}:
Experimentally, we couple a continuous-wave single-frequency laser through a single mode fiber into the cavity. An input image is loaded onto the DMD, which modifies the scattering potential in a reconfigurable manner. The entire modulation area of the DMD consists of 2560 $\times$ 1600 micromirrors with a pixel pitch of 7.6 $\mu$m. Each micromirror can be tilted by +15 or -15 degrees, representing the binary states +1 and -1. The input port of the cavity covers a portion of the DMD area (1260 $\times$ 784 micromirrors), which is exposed to light in the cavity; thus, we only modulate the micromirrors within this region. Instead of controlling individual micromirrors, we group micromirrors into macropixels, where all micromirrors in a single macropixel have the same tilt angle (binary state). To make images compatible for loading onto the DMD, we employed a binary thresholding method using the Floyd-Steinberg dithering algorithm   \cite{floyd1976}. 

We control the order of nonlinear mapping in our cavity in two ways, both changing the number of scattering events on the modulated area of the DMD. First, we reduce the DMD area where micromirros are toggled. Outside the modulated area, the micromirror configuration remains fixed. The number of bounces of light from a smaller modulated area is lower. By shrinking the dimension of macropixels by a factor of 4 or 16, total modulated area is reduced by the same factor. Alternatively, we can enhance the number of bounces with the DMD by increasing the dwell time of light inside the cavity. This is realized by covering the output port of the cavity with a partial reflector - a silicon wafer (thickness 0.63 mm), which partially reflects light at 1550 nm. As a result, the order of nonlinear random mapping increases.

The temporal coherent length of light exceeds the typical optical path length inside the cavity. The output light maintains high spatial coherence, resulting in a relatively high intensity contrast ($\sim$0.8) of output speckle pattern (after passing through a linear polarizer). Compared to the nonlinearity introduced by optical effects such as harmonic generation and self-phase modulation, a broadband pulsed laser is necessary to achieve the high pulse energy required for these nonlinear processes, producing much lower contrast. In addition, our system's nonlinear response is insensitive to optical power, more stable, and more energy-efficient. The output images recorded by our camera consistently display stable speckle patterns, with each speckle grain representing a distinct spatial mode. The number of output modes (speckle grains) in the camera image is determined by dividing the total area of the speckle grains used for computation by the average size of one speckle grain, which is derived from the full-width half-max of the intensity correlation function.\\

\noindent \textbf{FashionMNIST classification task}:
In the FashionMNIST \cite{xiao2017fashion} classification task, we study the impact of the number of modes and the modulated area size on DMD in terms of classification accuracy and mutual information between the output modes and the ground-truth target classes. To vary the number of modes, we crop the output camera image, controlling the number of output modes. This is achieved in \texttt{PyTorch} using the \texttt{nn.transform.CenterCrop} function. We manipulate the modulated area on the DMD by adjusting the macropixel size for the input data. For example, for the FashionMNIST dataset, we use a \(45\times28\) micro-pixels for each macropixelon the DMD, corresponding to one of the 28$\times$28 FashionMNIST image when we utilize the full modulated area. For a \(25\%\) modulated area, we use 22$\times$14 micropixels for one macropixel on the DMD. The entire set of 60,000 training data and 10,000 testing data are input sequentially on the DMD, and the corresponding camera speckle images are collected. We further applied a filter based on system stability (see Supplementary Section 2) to select images with a speckle stability over the threshold of 0.96. The data are then split in a 9:1 ratio to form training and testing datasets for classification. For classification, we employ ridge regressor from the \texttt{keras} package to train and infer with the output modes. Regarding the calculation of mutual information, detailed information on the algorithm can be found in Supplementary Section 4. We use the function \texttt{mutual\_info\_regression}, which takes vectors of pixel values in output modes and class labels, from the \texttt{feature selection} module in \texttt{scikit-learn}.\\

\noindent {\textbf{Programmable optical and digital parameters}
The maximum number of programmable optical parameters is given by the number of mirrors of the Digital Micromirror Device (DMD), which is approximately 4 million. The count of digitally programmable parameters, however, depends on the decoder utilized. Specifically, for the FashionMNIST classification task, the linear classifier requires only 1,000 parameters. In the task of FashionMNIST reconstruction, the parameter count increases to approximately 90k. For human face detection, the requirement is around 310k parameters, and for pedestrian tracking, the model uses around 1 million programmable parameters.}\\

\noindent \textbf{Training of digital decoder}:
For the tasks beyond classification, we start with low-dimensional vectors derived from the deep optical encoder --- the multiple-scattering cavity. Using these vectors, we train a digital decoder based on a neural network, with the objective of minimizing the mean squared loss relative to the ground truth targets' values in our training dataset. The dimensions for each target differed according to the tasks: 28$\times$28 for FashionMNIST image reconstruction, 15 sets of keypoints for human face keypoint detection, and four bounding box coordinates for pedestrian detection. The architecture selected for the decoding neural network is a multilayer perceptron, which incorporate batch normalization prior to each activation function. The ideal depths and widths of the hidden layers are determined by conducting a neural architecture search, initialized randomly at least 100 times to select the best architecture for the digital decoder. The same activation function, chosen from among \texttt{relu}, \texttt{tanh}, and \texttt{sigmoid}, is used during each search. All trainings are conducted on the NVIDIA A100 Tensor Core GPU via Google Colab. \\

\noindent \textbf{FashionMNIST image reconstruction task}:
In the FashionMNIST \cite{xiao2017fashion} image reconstruction task, we train a Multi-Layer Perceptron (MLP) as a digital decoder using pairs of 16 output modes (inputs) and ground truth FashionMNIST images (targets) to reconstruct the original images from the speckle patterns. To optimize the decoder's architecture, we employ neural architecture search, varying both depth and width to identify the best architecture for the decoder. We primarily study and compare two cases: case 1: Speckle features generated from a linear random projection through complex media, followed by quadratic detection on the optical field (to generate of linear optical features, we follow the methods in  \cite{saade2016random}); case 2: Speckle features generated from a nonlinear random mapping via a multiple-scattering cavity, again followed by quadratic detection on the optical field. In both scenarios, we ensure that the number of modes remains consistent, making the reconstruction quality comparable between the two cases. For the first case, the optimized decoder comprises a 2-layer MLP. For the second case, the optimized decoder utilizes a 4-layer MLP, both with the same activation function \texttt{sigmoid}. We further evaluate the reconstruction using a test dataset.\\

\noindent \textbf{Human face keypoint detection task}:
The human face keypoint detection dataset at Kaggle  \cite{facehuman} consists of facial keypoints, each characterized by a real-valued pair (x,y) indicating its position in the domain of pixel indices. This dataset identifies 15 specific keypoints corresponding to facial features, including centers of the left and right eyes, inner and outer corners of both eyes, inner and outer ends of both eyebrows, the tip of the nose, corners of the mouth on both sides, and the top and bottom centers of the lips. It's noteworthy to mention that the terms "left" and "right" are based on the subject's point of view. Some data points might not have all keypoints, which are represented as missing entries in the dataset. Each image in the dataset contains a list of pixels, with values ranging from 0 to 255, formatted for a resolution of 96x96 pixels. The training set includes 7049 images. Each row in this file provides the (x,y) coordinates of the 15 keypoints and image data in a row-ordered list of pixels. Conversely, the test set, comprises 1783 images; each row lists an ImageId and the corresponding row-ordered list of pixels for the image.

For data pre-processing, entries without keypoint information are excluded. In cases where an image had fewer than 15 keypoints, we duplicated some keypoints to ensure that all the labels consisted of 15 target points. This procedure ensures a consistent size of the MLP output layer.

Following this, the data are sent into a multiple-scattering cavity, with different modulated areas, corresponding to variable nonlinearity strengths, reminiscent of a deep neural networks encoding (see Supplementary Section 3). Only 16 output modes (using the \texttt{nn.transform.CenterCrop}) are extracted from this system. Using these modes, a digital decoder is developed based on neural architecture search, aiming to train on and infer the 15 facial keypoints.

Our analysis mainly compared two scenarios: one with approximately equal to 6.25$\%$ modulated area and another termed "100$\%$ $+$ enhanced", which is full modulated area bolstered by an extra partial reflector for enhanced scattering, as detailed in Supplementary Section 1. Our findings indicated that even with a decoder trained to its optimal capacity, the "100$\%$ $+$ enhanced" setup yielded better results. \\

\noindent \textbf{Pedestrian detection task}
In this task, we use the Caltech pedestrian dataset  \cite{dollar2009pedestrian}, one of the pioneering collections in the domain of computer vision, specifically designed for pedestrian detection tasks. This dataset has played an instrumental role in shaping the research trajectories in pedestrian detection, serving as a benchmark for numerous detection algorithms over the years. The data set offers a wide variety of real-world scenarios captured from urban settings, including pedestrians in various poses, occlusions, and varying light conditions. It provides an invaluable resource for the development and evaluation of algorithms, with its meticulous annotations and diverse challenges it poses.

Within this dataset, bounding boxes are utilized to accurately locate individual pedestrians in frames. These boxes are characterized by a set of four real-valued positions: $(x_1, y_1)$ for the top-left corner and $(x_2, y_2)$ for the bottom-right corner. Given the dynamic nature of urban environments, a single frame can contain multiple pedestrians, which results in multiple bounding boxes within that image.

To preprocess the dataset, we adopt a simplification strategy. Regardless of the number of bounding boxes present in the original image, we ensure that only one bounding box is retained per image. For images that contain multiple bounding boxes, only the first bounding box is selected and used as a label. In the cases where an image lacks a bounding box, it is removed from the dataset. All images from the Caltech dataset inherently possess 640$\times$480 resolutions. In our case, we downsampled the images to 320$\times$240 pixels. Our curated version of the dataset, divided into training and test segments, encapsulates a total of 10,000 images. 

Following the pre-processing steps, images are then introduced into a multiple-scattering cavity with full modulated area being enhanced by the partial reflector - the silicon wafer. From this system, a total of 25 output modes (using the \texttt{nn.transform.CenterCrop}) are derived. Harnessing these modes, we engineered a digital decoder rooted in the principles of neural architecture search. The overarching objective of this decoder is to train and subsequently infer the solitary bounding box in the images. 

We also generated Supplementary movies using the test dataset from various video locations. In these movies, the frame rate was reduced from the actual 30 fps to 9 fps for visualization purposes. The green boxes indicate the ground truth bounding boxes, while the magenta boxes represent the inferences. The actual inference time is well under 0.1 seconds. \\

\noindent \textbf{Data Availability Statement}: 
Partial example data pertaining to this study is available on GitHub at \url{https://github.com/comediaLKB/learning_with_passive_optical_nonlinear_mapping}. Additional data are available from the authors upon reasonable request. The training datasets used are publicly available at Fashion-MNIST \cite{xiao2017fashion}, the Human Face Keypoint Kaggle dataset \cite{facehuman}, and the Caltech Pedestrian dataset \cite{dollar2009pedestrian}.
\\

\noindent \textbf{Code Availability Statement}: 
Code is available on GitHub at \url{https://github.com/comediaLKB/learning_with_passive_optical_nonlinear_mapping}. 
\end{footnotesize}







\begin{thebibliography}{0}%
\makeatletter
\providecommand \@ifxundefined [1]{%
 \@ifx{#1\undefined}
}%
\providecommand \@ifnum [1]{%
 \ifnum #1\expandafter \@firstoftwo
 \else \expandafter \@secondoftwo
 \fi
}%
\providecommand \@ifx [1]{%
 \ifx #1\expandafter \@firstoftwo
 \else \expandafter \@secondoftwo
 \fi
}%
\providecommand \natexlab [1]{#1}%
\providecommand \enquote  [1]{``#1''}%
\providecommand \bibnamefont  [1]{#1}%
\providecommand \bibfnamefont [1]{#1}%
\providecommand \citenamefont [1]{#1}%
\providecommand \href@noop [0]{\@secondoftwo}%
\providecommand \href [0]{\begingroup \@sanitize@url \@href}%
\providecommand \@href[1]{\@@startlink{#1}\@@href}%
\providecommand \@@href[1]{\endgroup#1\@@endlink}%
\providecommand \@sanitize@url [0]{\catcode `\\12\catcode `\$12\catcode `\&12\catcode `\#12\catcode `\^12\catcode `\_12\catcode `\%12\relax}%
\providecommand \@@startlink[1]{}%
\providecommand \@@endlink[0]{}%
\providecommand \url  [0]{\begingroup\@sanitize@url \@url }%
\providecommand \@url [1]{\endgroup\@href {#1}{\urlprefix }}%
\providecommand \urlprefix  [0]{URL }%
\providecommand \Eprint [0]{\href }%
\providecommand \doibase [0]{https://doi.org/}%
\providecommand \selectlanguage [0]{\@gobble}%
\providecommand \bibinfo  [0]{\@secondoftwo}%
\providecommand \bibfield  [0]{\@secondoftwo}%
\providecommand \translation [1]{[#1]}%
\providecommand \BibitemOpen [0]{}%
\providecommand \bibitemStop [0]{}%
\providecommand \bibitemNoStop [0]{.\EOS\space}%
\providecommand \EOS [0]{\spacefactor3000\relax}%
\providecommand \BibitemShut  [1]{\csname bibitem#1\endcsname}%
\let\auto@bib@innerbib\@empty
\end{thebibliography}%


\section*{References}
\begin{thebibliography}{99}
\bibitem{shastri2018}
Prucnal, Paul R., and Bhavin J. Shastri. Neuromorphic photonics. CRC press, 2017.

\bibitem{kues2017}
Kues, M., Reimer, C., Roztocki, P., Cortes, L.R., Sciara, S., Wetzel, B., Zhang, Y., Cino, A., Chu, S.T., Little, B.E., et al. (2017). On-chip generation of high-dimensional entangled quantum states and their coherent control. \textit{Nature}, 546(7660), 622--626.

\bibitem{xu202111}
Xu, X., Tan, M., Corcoran, B., Wu, J., Boes, A., Nguyen, T.G., Chu, S.T., Little, B.E., Hicks, D.G., Morandotti, R., et al. (2021). 11 TOPS photonic convolutional accelerator for optical neural networks. \textit{Nature}, 589(7840), 44--51. Nature Publishing Group UK London.

\bibitem{wetzstein2020inference}
Wetzstein, G., Ozcan, A., Gigan, S., Fan, S., Englund, D., Soljačić, M., Denz, C., Miller, D.A.B., $\&$ Psaltis, D. (2020). Inference in artificial intelligence with deep optics and photonics. \textit{Nature}, 588(7836), 39--47. Nature Publishing Group UK London.

\bibitem{shastri2021photonics}
Shastri, B.J., Tait, A.N., Ferreira de Lima, T., Pernice, W.H.P., Bhaskaran, H., Wright, C.D., $\&$ Prucnal, P.R. (2021). Photonics for artificial intelligence and neuromorphic computing. \textit{Nature Photonics}, 15(2), 102--114. Nature Publishing Group UK London.

\bibitem{shen2017deep}
Shen, Y., Harris, N.C., Skirlo, S., Prabhu, M., Baehr-Jones, T., Hochberg, M., Sun, X., Zhao, S., Larochelle, H., Englund, D., et al. (2017). Deep learning with coherent nanophotonic circuits. \textit{Nature Photonics}, 11(7), 441--446.

\bibitem{rotter2017light}
Rotter, S., $\&$ Gigan, S. (2017). Light fields in complex media: Mesoscopic scattering meets wave control. \textit{Reviews of Modern Physics}, 89(1), 015005.

\bibitem{chang2018hybrid}
Chang, J., Sitzmann, V., Dun, X., Heidrich, W., $\&$ Wetzstein, G. (2018). Hybrid optical-electronic convolutional neural networks with optimized diffractive optics for image classification. \textit{Scientific Reports}, 8(1), 12324.

\bibitem{hughes2018training}
Hughes, T.W., Minkov, M., Shi, Y., $\&$ Fan, S. (2018). Training of photonic neural networks through in situ backpropagation and gradient measurement. \textit{Optica}, 5(7), 864--871. Optica Publishing Group.

\bibitem{goodfellow2016deep}
Goodfellow, I., Bengio, Y., $\&$ Courville, A. (2016). \textit{Deep learning}. MIT Press.

\bibitem{lecun2015deep}
LeCun, Y., Bengio, Y., $\&$ Hinton, G. (2015). Deep learning. \textit{Nature}, 521(7553), 436--444.

\bibitem{wang2023image}
Wang, T., Sohoni, M.M., Wright, L.G., Stein, M.M., Ma, S.-Y., Onodera, T., Anderson, M.G., $\&$ McMahon, P.L. (2023). Image sensing with multilayer nonlinear optical neural networks. \textit{Nature Photonics}, 17(5), 408--415. Nature Publishing Group UK London.

\bibitem{krizhevsky2012imagenet}
Krizhevsky, A., Sutskever, I., $\&$ Hinton, G.E. (2012). Imagenet classification with deep convolutional neural networks. \textit{Advances in Neural Information Processing Systems}, 25.

\bibitem{lin2018all}
Lin, X., Rivenson, Y., Yardimci, N.T., Veli, M., Luo, Y., Jarrahi, M., $\&$ Ozcan, A. (2018). All-optical machine learning using diffractive deep neural networks. \textit{Science}, 361(6407), 1004--1008. American Association for the Advancement of Science.

\bibitem{tait2017neuromorphic}
Tait, A.N., De Lima, T.F., Zhou, E., Wu, A.X., Nahmias, M.A., Shastri, B.J., $\&$ Prucnal, P.R. (2017). Neuromorphic photonic networks using silicon photonic weight banks. \textit{Scientific Reports}, 7(1), 7430. Nature Publishing Group UK London.

\bibitem{miller2015are}
Miller, D.A.B. (2015). Are optical transistors the logical next step? \textit{Nature Photonics}, 9(1), 10--13. Nature Publishing Group.

\bibitem{wang2018chip}
Wang, M.M., Pagani, M., $\&$ Eggleton, B.J. (2018). A chip-integrated coherent photonic-phononic memory. \textit{Nature Communications}, 9(1), 1--7. Nature Publishing Group.

\bibitem{teugin2021scalable}
Teğin, U., Yıldırım, M., Oğuz, İ., Moser, C., $\&$ Psaltis, D. (2021). Scalable optical learning operator. \textit{Nature Computational Science}, 1(8), 542--549. Nature Publishing Group US New York.

\bibitem{williamson2019reprogrammable}
Williamson, I. A. D., Hughes, T. W., Minkov, M., Bartlett, B., Pai, S., $\&$ Fan, S. (2019). Reprogrammable electro-optic nonlinear activation functions for optical neural networks. \textit{IEEE Journal of Selected Topics in Quantum Electronics}, 26(1), 1--12. IEEE.


\bibitem{li2022all}
Li, G.H.Y., Sekine, R., Nehra, R., Gray, R.M., Ledezma, L., Guo, Q., $\&$ Marandi, A. (2022). All-optical ultrafast ReLU function for energy-efficient nanophotonic deep learning. \textit{Nanophotonics}, 12(5), 847--855. De Gruyter.

\bibitem{zhou2022nonlinear}
Zhou, T., Scalzo, F., $\&$ Jalali, B. (2022). Nonlinear Schrödinger kernel for hardware acceleration of machine learning. \textit{Journal of Lightwave Technology}, 40(5), 1308--1319. IEEE.

\bibitem{shirdel2016photonic}
Shirdel, M., $\&$ Mansouri-Birjandi, M.A. (2016). Photonic crystal all-optical switch based on a nonlinear cavity. \textit{Optik}, 127(8), 3955--3958. Elsevier.

\bibitem{eliezer2022exploiting}
Eliezer, Y., Ruhrmair, U., Wisiol, N., Bittner, S., $\&$ Cao, H. (2023). Tunable nonlinear optical mapping in a multiple-scattering cavity. \textit{Proceedings of the National Academy of Sciences}, 120(31), e2305027120.

\bibitem{saade2016random}
Saade, A., Caltagirone, F., Carron, I., Daudet, L., Drémeau, A., Gigan, S., $\&$ Krzakala, F. (2016). Random projections through multiple optical scattering: Approximating kernels at the speed of light. In \textit{2016 IEEE International Conference on Acoustics, Speech and Signal Processing (ICASSP)} (pp. 6215--6219). IEEE.

\bibitem{rafayelyan2020large}
Rafayelyan, M., Dong, J., Tan, Y., Krzakala, F., $\&$ Gigan, S. (2020). Large-scale optical reservoir computing for spatiotemporal chaotic systems prediction. \textit{Physical Review X}, 10(4), 041037.

\bibitem{dong2019optical}
Dong, J., Rafayelyan, M., Krzakala, F., $\&$ Gigan, S. (2019). Optical reservoir computing using multiple light scattering for chaotic systems prediction. \textit{IEEE Journal of Selected Topics in Quantum Electronics}, 26(1), 1--12.

\bibitem{brossollet2021lighton}
Brossollet, C., Cappelli, A., Carron, I., Chaintoutis, C., Chatelain, A., Daudet, L., Gigan, S., Hesslow, D., Krzakala, F., Launay, J., $\&$ others. (2021). LightOn Optical Processing Unit: Scaling-up AI and HPC with a Non von Neumann co-processor. In \textit{2021 IEEE Hot Chips 33 Symposium (HCS)} (pp. 1--11). IEEE.

\bibitem{ohana2021photonic}
Ohana, R., Medina, H., Launay, J., Cappelli, A., Poli, I., Ralaivola, L., $\&$ Rakotomamonjy, A. (2021). Photonic differential privacy with direct feedback alignment. \textit{Advances in Neural Information Processing Systems}, 34, 22010--22020.


\bibitem{agrawal2000nonlinear}
Agrawal, G.P. (2000). Nonlinear fiber optics. In \textit{Nonlinear Science at the Dawn of the 21st Century} (pp. 195--211). Springer.

\bibitem{boyd2008nonlinear}
Boyd, R. W., Gaeta, A. L., $\&$ Giese, E. (2008). Nonlinear optics. In \textit{Springer Handbook of Atomic, Molecular, and Optical Physics} (pp. 1097--1110). Springer.


\bibitem{wang2016thermo}
Wang, J., Zhu, B., Hao, Z., Bo, F., Wang, X., Gao, F., Li, Y., Zhang, G., $\&$ Xu, J. (2016). Thermo-optic effects in on-chip lithium niobate microdisk resonators. \textit{Optics Express}, 24(19), 21869--21879. Optica Publishing Group.


\bibitem{ryou2021free}
Ryou, A., Whitehead, J., Zhelyeznyakov, M., Anderson, P., Keskin, C., Bajcsy, M., $\&$ Majumdar, A. (2021). Free-space optical neural network based on thermal atomic nonlinearity. \textit{Photonics Research}, 9(4), B128--B134. Optica Publishing Group.


\bibitem{ohtsubo2013}
Ohtsubo, J. (2013). Semiconductor Lasers Stability, Instability and Chaos Second, Enlarged Edition. \textit{SPRINGER SERIES IN OPTICAL SCIENCES}, 111. SPRINGER.

\bibitem{xiao2017fashion}
Xiao, H., Rasul, K., $\&$ Vollgraf, R. (2017). Fashion-mnist: a novel image dataset for benchmarking machine learning algorithms. \textit{arXiv preprint arXiv:1708.07747}.

\bibitem{cover2012elements}
Cover, T. M., $\&$ Thomas, J. A. (2012). Elements of information theory. \textit{Wiley}.


\bibitem{donoho2006compressed}
Donoho, D. L. (2006). Compressed sensing. \textit{IEEE Transactions on Information Theory}, 52(4), 1289--1306. IEEE.

\bibitem{peng2005feature}
Peng, H., Long, F., $\&$ Ding, C. (2005). Feature selection based on mutual information criteria of max-dependency, max-relevance, and min-redundancy. \textit{IEEE Transactions on Pattern Analysis and Machine Intelligence}, 27(8), 1226--1238. IEEE.

\bibitem{chen2016xgboost}
Chen, T., $\&$ Guestrin, C. (2016). Xgboost: A scalable tree boosting system. In \textit{Proceedings of the 22nd ACM SIGKDD International Conference on Knowledge Discovery and Data Mining} (pp. 785--794).

\bibitem{kraskov2004estimating}
Kraskov, A., St{\"o}gbauer, H., $\&$ Grassberger, P. (2004). Estimating mutual information. \textit{Physical Review E}, 69(6), 066138. APS.

\bibitem{facehuman}
Petterson, J., $\&$ Cukierski, W. (2013). Facial Keypoints Detection Dataset (provided by Dr. Yoshua Bengio of the University of Montreal). Kaggle. Retrieved from https://kaggle.com/competitions/facial-keypoints-detection


\bibitem{dollar2009pedestrian}
Doll{\'a}r, P., Wojek, C., Schiele, B., $\&$ Perona, P. (2009). Pedestrian detection: A benchmark. In \textit{2009 IEEE Conference on Computer Vision and Pattern Recognition} (pp. 304--311). IEEE.

\bibitem{skalli2022photonic}
Skalli, A., Robertson, J., Owen-Newns, D., Hejda, M., Porte, X., Reitzenstein, S., Hurtado, A., $\&$ Brunner, D. (2022). Photonic neuromorphic computing using vertical cavity semiconductor lasers. \textit{Optical Materials Express}, 12(6), 2395--2414. Optica Publishing Group.

\bibitem{van2017advances}
Van der Sande, G., Brunner, D., $\&$ Soriano, M. C. (2017). Advances in photonic reservoir computing. \textit{Nanophotonics}, 6(3), 561--576. De Gruyter.

\bibitem{boikov2023direct}
Boikov, I. K., Brunner, D., $\&$ De Rossi, A. (2023). Evanescent coupling of nonlinear integrated cavities for all-optical reservoir computing. \textit{New Journal of Physics}, 25(9), 093056. IOP Publishing.

\bibitem{porte2021complete}
Porte, X., Skalli, A., Haghighi, N., Reitzenstein, S., Lott, J. A., $\&$ Brunner, D. (2021). A complete, parallel and autonomous photonic neural network in a semiconductor multimode laser. \textit{Journal of Physics: Photonics}, 3(2), 024017. IOP Publishing.

\bibitem{wright2022deep}
Wright, L. G., Onodera, T., Stein, M. M., Wang, T., Schachter, D. T., Hu, Z., $\&$ McMahon, P. L. (2022). Deep physical neural networks trained with backpropagation. \textit{Nature}, 601(7894), 549--555. Nature Publishing Group UK London.

\bibitem{chen2023deep}
Chen, Z., Sludds, A., Davis III, R., Christen, I., Bernstein, L., Ateshian, L., Heuser, T., Heermeier, N., Lott, J. A., Reitzenstein, S., $\&$ others. (2023). Deep learning with coherent VCSEL neural networks. \textit{Nature Photonics}, 1--8. Nature Publishing Group UK London.

\bibitem{momeni2023physics}
Momeni, A., Guo, X., Lissek, H., $\&$ Fleury, R. (2023). Physics-inspired Neuroacoustic Computing Based on Tunable Nonlinear Multiple-scattering. \textit{arXiv preprint arXiv:2304.08380}.

\bibitem{del2018leveraging}
del Hougne, P., $\&$ Lerosey, G. (2018). Leveraging chaos for wave-based analog computation: Demonstration with indoor wireless communication signals. \textit{Physical Review X}, 8(4), 041037.

\bibitem{momeni2023backpropagation}
Momeni, A., Rahmani, B., Malléjac, M., Del Hougne, P., $\&$ Fleury, R. (2023). Backpropagation-free training of deep physical neural networks. \textit{Science}, 382(6676), 1297--1303.


\bibitem{devlin2018bert}
Devlin, J., Chang, M.-W., Lee, K., $\&$ Toutanova, K. (2019). Bert: Pre-training of deep bidirectional transformers for language understanding. \textit{Proceedings of NAACL-HLT}.

\bibitem{chen2015optical}
Chen, C. L., Mahjoubfar, A., $\&$ Jalali, B. (2015). Optical data compression in time stretch imaging. \textit{PloS one}, \textit{10}(4), e0125106.


\bibitem{weng2023non}
Weng, X., Feng, J., Perry, A., $\&$ Vuong, L.T. (2023). Non-Line-of-Sight Full-Stokes Polarimetric Imaging with Solution-Processed Metagratings and Shallow Neural Networks. \textit{ACS Photonics}.

\bibitem{li2021spectrally}
Li, J., Mengu, D., Yardimci, N.T., Luo, Y., Li, X., Veli, M., Rivenson, Y., Jarrahi, M., $\&$ Ozcan, A. (2021). Spectrally encoded single-pixel machine vision using diffractive networks. \textit{Science Advances}, 7(13), eabd7690.


\bibitem{muminov2020fourier}
Muminov, B., $\&$ Vuong, L.T. (2020). Fourier optical preprocessing in lieu of deep learning. \textit{Optica}, 7(9), 1079--1088.


\bibitem{wu2014secure}
Wu, B., Shastri, B.J., $\&$ Prucnal, P.R. (2014). Secure communication in fiber-optic networks. In \textit{Emerging trends in ICT security} (pp. 173--183). Elsevier.

\bibitem{venkataraman2013phase}
Venkataraman, V., Saha, K., $\&$ Gaeta, A.L. (2013). Phase modulation at the few-photon level for weak-nonlinearity-based quantum computing. \textit{Nature Photonics}, 7(2), 138--141.


\bibitem{bezzam2022privacy}
Bezzam, E., Vetterli, M., $\&$ Simeoni, M. (2022). Privacy-enhancing optical embeddings for lensless classification. \textit{arXiv preprint arXiv:2211.12864}.


\bibitem{cappelli2022adversarial}
Cappelli, A., Ohana, R., Launay, J., Meunier, L., Poli, I., $\&$ Krzakala, F. (2022). Adversarial robustness by design through analog computing and synthetic gradients. In \textit{ICASSP 2022-2022 IEEE International Conference on Acoustics, Speech and Signal Processing (ICASSP)} (pp. 3493--3497). IEEE.

\bibitem{yildirim2023nonlinear}
Yildirim, M., Dinc, N. U., Oguz, I., Psaltis, D., $\&$ Moser, C. (2024). Nonlinear Processing with Linear Optics. \textit{Nature Photonics}, \textit{18}, XXX-XXX.

\bibitem{wanjura2023fully}
Wanjura, C.C., $\&$ Marquardt, F. (2023). Fully Non-Linear Neuromorphic Computing with Linear Wave Scattering. \textit{arXiv preprint arXiv:2308.16181}.

\bibitem{floyd1976}
Floyd, R. W., $\&$ Steinberg, L. (1976). An Adaptive Algorithm for Spatial Greyscale. In \textit{Proceedings of the Society for Information Display} (pp. 36--37).

\bibitem{lecun1998mnist}
LeCun, Y. (1998). The MNIST database of handwritten digits. http://yann. lecun. com/exdb/mnist/.

\end{thebibliography}
\end{document}